**The central role of thermal collective strain in the relaxation of structure in a supercooled liquid**

Asaph Widmer-Cooper[1,2] and Peter Harrowell[1]

[1]School of Chemistry, University of Sydney, Sydney New South Wales 2006 Australia

and

[2]Materials Science Division, Lawrence Berkeley National Laboratory, Berkeley, California 94720

Abstract

The spatial distribution of structural relaxation in a supercooled liquid is studied using molecular dynamics simulations of a 2D binary mixture. It is shown that the spatial heterogeneity of the relaxation along with the time scale of the relaxation is determined, not by the frequency with which particles move a distance $\pi/2k_{Bragg}$, but by the frequency with which particles can achieve persistent displacements. We show that these persistent displacements are achieved through the coupled action of local reorganizations and unrecoverable thermal strains.

Unlike the diffusion constant, the structural relaxation time scales with temperature in a manner similar to the shear viscosity at large supercoolings [1]. As it is the large increase of the shear viscosity on supercooling that defines the glass transition, understanding the associated structural relaxation is an important issue. What kinds of particle motion are responsible for structural relaxation in a dense amorphous material? In this paper we present evidence from molecular dynamics simulations to show that structural relaxation is associated with the heterogeneous distribution, not of the *amplitude* of particle displacements, but of the *irreversibility* of particle motion (identified here by the lifetime of particle displacements away from their original positions). The slow down of structural relaxation on cooling results from the associated scarcity of these regions of persistent displacement. Moving beyond the simple language of 'cage escapes', we show that the persistent displacements responsible for structural relaxation consist of long-lived strain-like motions associated with one or more sites of more extensive reorganization.

To begin we require a measure of structural relaxation in which each particle's contribution to structural memory is explicit. To this end, we define the following structural relaxation function. Let $F_d(t)$ be defined as

$$F_d(t) = \frac{1}{N}\left\langle \sum_i w_i(d,t) \right\rangle \qquad (1)$$

where $w_i(d,t) = 1$ if the particle is within a distance $d$ of its initial position at a time $t$ and zero otherwise. The average is over the initial time. We shall choose $d = \pi/2k_{Bragg}$, where $k_{Bragg}$ is the magnitude of the wave vector of the first maximum in the total structure factor $S(k)$. This value of $d$ is the shortest distance for which $\mathrm{Re}[\exp(-ik_{Bragg}d)] = 0$. We shall refer to this volume about a particle's initial site as the 'cell'.

Functions similar or identical to $F_d(t)$ have been used previously to study the four-point susceptibility in MD simulations [2] and granular material [3,4]. All of these groups examined how the choice of $d$ affected their correlation functions and chose values of $d$ so as to achieve a maximum [2,4] or near maximum [3] in the height of the $\chi_4(t)$ peak. In contrast, we treat $F_d(t)$ as a structural relaxation function and hence, as in the case of the intermediate scattering function, the length scale $d$ is fixed by the choice of the wave vector associated with that structure. The value of $d$ used here ($d = 0.283$) to study structural relaxation is similar to or somewhat smaller than the values 0.3 [2] and 0.5 [3] used in previous studies. While $F_d(t)$ monitors only self motion there is considerable evidence [5,6] supporting the proposition that in the supercooled liquid the individual particle motions have become so tightly correlated with the collective motion of the surrounding particles that there is little difference between the relaxation behaviour of the self and collective correlation functions.

For a glass-forming liquid, we use a two-dimensional (2D) equimolar binary mixture of particles interacting via purely repulsive potentials of the form

$$u_{ab}(r) = \varepsilon \left[ \frac{\sigma_{ab}}{r} \right]^{12} \qquad (2)$$

where $\sigma_{12} = 1.2 \times \sigma_{11}$ and $\sigma_{22} = 1.4 \times \sigma_{11}$. All units quoted will be reduced so that $\sigma_{11} = \varepsilon = m = 1.0$ where $m$ is the mass of both types of particle. Specifically, the reduced unit of time is given by $\tau = \sigma_{11} (m/\varepsilon)^{1/2}$ so that, at $T = 0.4$, the average time between velocity reversals is $0.1\tau$ [1]. A total of $N = 1024$ particles were enclosed in a square box with periodic boundary conditions. We note that 2D liquids are playing an increasingly important role in resolving the complex dynamics near the glass transitions in both experiments [7,8] and simulations [9]. This model and its approach to the glass transition have been studied in detail and readers are directed to these papers [10] for more information. The simulations were carried out at constant pressure ($P = 13.5$) using a Nosé-Poincaré-Andersen algorithm [11]. We note that the rescaling of particle positions associated with the volume fluctuations of the $NPT$ ensemble can contribute to the rate at which particles escape their cell. We have checked for this by repeating the calculations in the $NVT$ ensemble at $T = 0.4$ and find no significant difference in the distribution of escape rates or lifetimes.

In Figs. 1 and 2 we compare the time dependence of the relaxation functions and the temperature dependence of the relaxation times obtained from $F_d(t)$, where $d = 0.283\sigma_{11}$, and from the intermediate scattering function $F_s(k_2,t)$, the latter function referring to the relaxation of the large particle subpopulation. (The relaxation time is defined as the time at which the relaxation function equals $1/e$. In the case of the intermediate scattering function this time is called $\tau_\alpha$.) We find that the new structural relaxation function $F_d(t)$ closely resembles the intermediate scattering function $F_s(k_2,t)$ [7], and that the relaxation times obtained from the two functions exhibit almost identical magnitudes and temperature dependence. In Fig. 3 we show a sequence of maps, taken from a single trajectory at $T = 0.4$, showing the spatial distribution of particles that have left their cells as relaxation proceeds. While there are considerable fluctuations between snapshots in Fig. 3, it is evident that there is an underlying spatial heterogeneity associated with structural relaxation. In this paper we shall establish the source of this heterogeneity.

Our goal in this paper is to describe the particle motions responsible for structural relaxation. An obvious (and, we shall see, incorrect) approach would be to look at how each particle first escaped from its cell. The problem with this approach is evident in Fig. 4 where we plot the fraction of particles at $T = 0.4$ that have managed to remain outside their cell for a period longer than $x$ during the time interval $t$. Selecting $x = 0$ (i.e. counting any escape at all, no matter how short-lived) we see that the fraction of particles that have never left their cell decays very rapidly with time $t$, much faster than the structural relaxation itself. This is because escapes are both widespread and, overwhelmingly, quickly reversed. The length scale associated with structural relaxation (which, for our purposes, is the radius of the cell) is simply small enough that particles are capable of departing without significant local rearrangements. (Even in the single component small particle crystal at $T = 0.4$, well below $T_m = 0.95$, roughly 60% of the particles have escaped their initial cells at least once within $10\tau$.)

Since most departures of a particle from its cell are reversed, the escape events cannot, in general, be directly associated with irreversible relaxation. What we are interested in are those rarer events where the movement out of the cell occurs in such a way as to be unrecoverable over some time interval, sufficiently long to filter out reversible vibrations but not so long as to see the motion corresponding to the elementary processes of relaxation being 'over-written' by subsequent particle movements. This problem, the identification an elementary irreversible event, is similar to that addressed by Heuer and coworkers [12] in implementing the idea of metabasins. The difference between this work and that of ref. [12] is that here we need to identify the real space character of the elementary irreversible processes. We believe that a time interval in the range 10-20$\tau$ satisfies these requirements. (For comparison, the transverse wave at $T = 0.4$ traverses the sample in $\sim 7\tau$ [13].) This conclusion is based on the following three observations:

1) In Fig. 4 we have plotted the time dependence at $T = 0.4$ of the fraction of particles that have not yet managed to achieve an escape lifetime of $x$, where $x = 0, 1, 10, 20$ and $50\tau$. The decay of each of these functions represents the rate at which particles are achieving escapes of various life times. If we compare these decays with that of the structural relaxation function we see that around the time at which $F_d \sim 1/e$, the decay curves for $x = 10\tau$ and $20\tau$ straddle the structural relaxation function.

2) In Fig. 5 we have plotted a time ordered sequence of maps showing the positions of particles that have managed an escape of duration $x$ at least once during the time of observation the value of $x$ is indicated by different colours.  It is clearly evident from these maps that heterogeneity in structural relaxation arises, not from the act of cell escape itself, which quickly achieves a homogeneous distribution, but from spatial variation in the *duration* of these escapes.

3) The spatial correlations among those particles that are reluctant to return to their cells, evident in Fig. 5, can be quantified in terms of cluster analysis. In Fig. 6 we show how the maximum size of clusters comprised of the first 10% of particles to have escape lifetime in excess of $x$ depends on the choice of $x$ for a range of temperatures.  We find that that the clustering reaches a maximum at around an escape lifetime of $x = 10\tau$ for all $T \leq 1$. (The distribution of the average cluster size shows a similar maximum at an escape lifetime of $10\tau$.) The clustering of the particles involved in persistent displacement increases significantly as the temperature is lowered.

Having settled on an escape lifetime between 10-20τ as defining the 'elementary' processes of relaxation, we can return to our goal, the physical description of the particle movements that constitute these processes. Given the possible complexity of the geometry of the motions, we shall focus on the topological changes only. If the motion resembled the 'cage escape' frequently mentioned in the context of supercooled liquids, we would expect the escapee to have quickly lost a significant fraction of the neighbours that it had immediately prior to its departure from the cell. Conversely, if the motion resembled a strain deformation, then a particle could escape its cell while still retaining all of these initial neighbours. We shall refer to a displaced particle that has lost no more than one neighbour as having been involved in a *strain*, other displacements will be referred to as *reorganizations*. In Fig. 7 we have plotted the fraction of escaped particles that are identified as being involved in a strain as a function of the time elapsed since their escape. We find that roughly 80-90% of particles escape their cell as part of a strain-like motion and, after an escape lifetime of 10τ, from 30% ($T = 1$) to 70% ($T = 0.5$) of the escaped particles still remain surrounded by all or almost all of their initial neighbours. These observations lead us to our second significant result – that a substantial fraction of particles that made a long-lived contribution to structural relaxation did so, not by some significant local rearrangement but, rather, by thermally generated unrecoverable strains. As shown in Fig. 7, these strains can persist as strains for a substantial fraction of the structural memory time of the liquid (e.g. of the escapes at $T = 0.4$ that persist for $\tau_\alpha$, 30% persist as strains).

In Fig. 8, we present an initial configuration at $T = 0.4$ with those particles that were involved in strains or reorganizations when their escape lifetime reached 20τ indicated by open and filled circles, respectively. The particles involved in strains make up compact extended domains while those particles associated with reorganizations appear in small groups or in isolation. The colour coding reveals the groups of particles that left their cells at roughly similar times. The presence of well-defined domains of roughly coincident motions in which neighbourhoods were retained is consistent with our description of the motion as a collective strain. Significantly, we also see that reorganization events typically occur at similar times and locations to the local strain events. This suggests that the two are coupled. Such a coupling could certainly account for why the strains were irreversible. The complex reorganization event would frustrate not only its own reversal but also that of the strain field coupled to it. In Fig. 9 we plot the analogous time correlated maps as in Fig. 8 but for a larger system with $N = 4096$ particles. We observe the same temporal and spatial correlations between the localised reorganizations and unrecoverable strains as seen in the smaller system

Our two conclusions are: (i) that the rate of structural relaxation and its spatial heterogeneity are determined by the statistics of localized variation, not in the amplitude of particle displacements, but in the probability that these small displacements (of size $\pi/2k_{Bragg}$) are long lived, and (ii) that these long lived displacements take the form of thermally excited collective strains in association with highly localised particle reorganizations. The heterogeneities relevant to structural relaxation, in other words, are not to be defined in terms of slow/fast or immobile/mobile particles but correspond, instead, to irreversible/reversible displacements. These conclusions lead us on to two fundamental questions: what causes the localization and what causes irreversibility? Previously, we have shown that the spatial distribution of reorganizations, as quantified by the loss of four nearest neighbours, correlates well with the spatial distribution of soft quasi-localized modes [14]. While we have yet to complete the comparison between the modes and the irreversible motions identified in this study, it seems reasonable to suggest that these modes are also responsible for the heterogeneity we see in the structural relaxation.

Explaining irreversibility is a more awkward proposition. Irreversibility is a consequence of dynamics accessing a sufficient volume of configuration space such that the trajectory can 'get lost' and, as a result, have a low probability of returning. This idea represents the kinetic (as opposed to thermodynamic) justification behind the introduction of the entropic droplet model [15] for relaxation in the amorphous energy landscape. While a collective strain may not involve a sufficient number of independent degrees of freedom to accomplish irreversibility, reorganizations (typically) do. Our results suggest a picture in which structural relaxation is achieved by the combined action of reorganizations and the extended strains coupled to them and rendered irreversible by virtue of this coupling. Experimental visualizations of relaxation in 2D granular media [8] appear to support our conclusions concerning the importance of strain-like motions in structural relaxation.

Whereas we have considered relaxation in a liquid at equilibrium (albeit a metastable one), there has been considerable work on structural transitions in non-equilibrium amorphous phases undergoing shear [16-24]. In 1979 Argon and Kuo [16] identified an archetypical reorganization event involved in plastic shear flow, a rotation of a pair of particles, which they called a shear transformation zone (STZ). More recently, analysis of materials under shear have identified local processes related to plastic behaviour by determining the degree to which particle displacements deviate from that expected for an affine deformation [22]. The concept of non-affine deformations generalises the microscopic description of plastic behaviour to explicitly include strain-like contributions to plastic flow along with reorganizations such as the STZ's. At zero temperature ('quasi-static'), one can ask whether the displacements resulting from an applied shear strain can be reversed with the reversal of the strain. Applying this mechanical criterion, Lundberg et al [23] have shown that STZ events can be either irreversible or reversible. Finally, a number of recent studies [24] have described the extended strain fields associated with relaxing mechanically induced stresses under conditions of plastic flow. This body of work contains clear parallels with the results described in this paper for the relaxation of thermal fluctuations at equilibrium. Exploring the relationship between these two pictures of amorphous relaxation - on one hand, the equilibrium one of soft local modes, and thermal reorganizations coupled to strain fields, and, on the other hand, the non-equilibrium quasi-static picture of STZ's and non-affine strain fields – holds the promise of unifying the spatial description and, hence, physical understanding of the microscopic mechanisms of structural and stress relaxation in disordered materials.

**FIGURE CAPTIONS**

**Figure 1.** The relaxation functions $F_d(t)$ and $F_s(k_2,t)$ (with $k_2 = 5.36\sigma_{11}^{-1}$) at $T =$ (from left to right) 1, 0.8, 0.6, 0.55, 0.5, 0.46, and 0.4.

**Figure 2**. Arrhenius plots of the relaxation times $\tau_e$ from $F_d(t)$ and $F_s(k_2,t)$ (with $k_2 = 5.36\sigma_{11}^{-1}$) as a function of temperature.

**Figure 3.** The spatial heterogeneity of structural relaxation. A sequence of configurations generated during a $T = 0.4$ trajectory in which particles are indicated as filled circles if, at each of the indicated times, they are outside of their initial cells.

**Figure 4.** The fraction $F_{cell}(x,t)$ of particles at $T = 0.4$ that have not yet left their cells for a period of time greater than $x$, where $x = 0, 1, 10, 20$ and $50\tau$, as a function of $t$ the time of observation. For comparison, the function $F_d(t)$, the fraction of particles in their cells after time $t$, is also plotted for $T = 0.4$. A dashed line indicates $F_d(t) = 1/e$.

**Figure 5.** A sequence of configurations generated during a $T = 0.4$ trajectory (the same as used in Fig. 2). Particles are colored according to their maximum escape lifetime (in units of $\tau$) during the elapsed time $t$.

**Figure 6.** The maximum size of cluster formed from the first 10% of particles to achieve an escape lifetime $x$ as a function of $x$ for $T = 0.4, 0.46, 0.5, 0.6$ and 1. Note that the peak occurs around $10\tau$ for all temperatures. The error bars represent one standard error.

**Figure 7.** The fraction of escaped particles that are involved in strains $F_{strain}(t)$ as a function of the time $t$ since their escape. Data is shown for $T = 0.4$, 0.46, 0.5, 0.6 and 1. The vertical dotted line indicates the 'persistence' time $10\tau$.

**Figure 8**. An initial configuration at $T = 0.4$ is shown in which each particle is depicted as either a filled or open circle depending on whether they were involved in a reorganization or a strain, respectively, when their escape lifetime reached $20\tau$. The color scale indicates the time (in $\tau$) during the trajectory at which each particle achieved its $20\tau$ escape lifetime. Particles with the same colour reached their $20\tau$ lifetime within the same time period.

**Figure 9**. An initial configuration of a 4096 particle system at $T = 0.4$. Each particle is depicted as either a filled or open circle depending on whether they were involved in a reorganization or a strain, respectively, when their escape lifetime reached $20\tau$. Particles with the same color reached this lifetime within the same time period. Large and small circles indicate large and small particles.

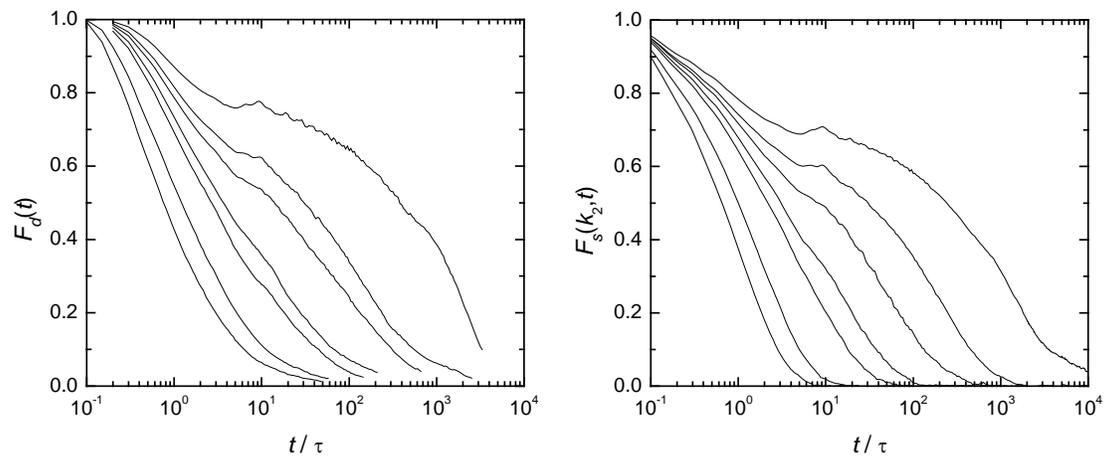

Figure 1

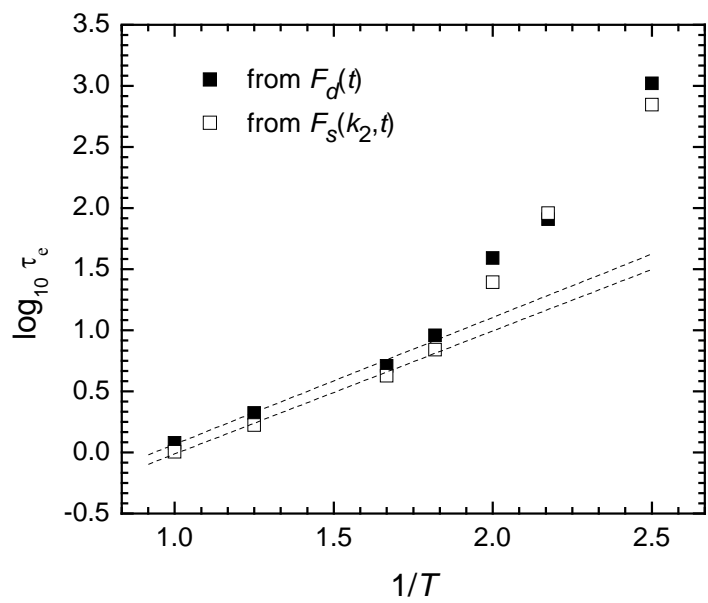

Figure 2

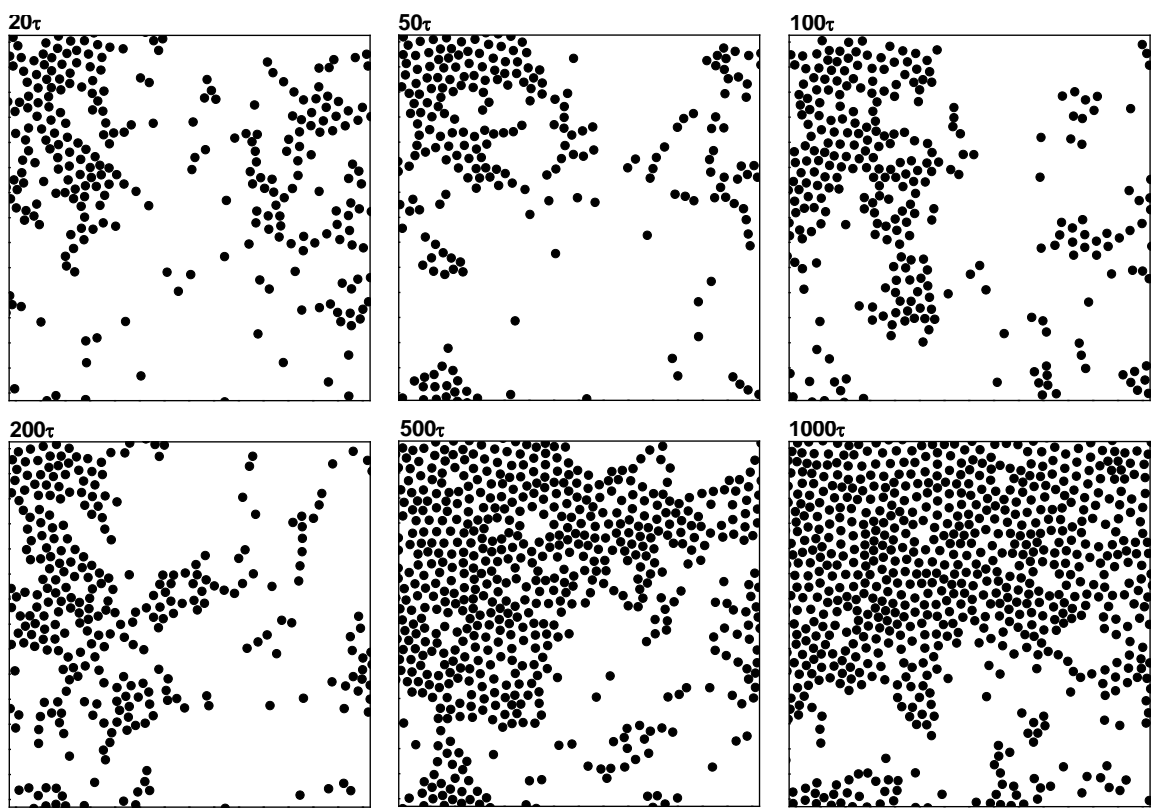

Figure 3

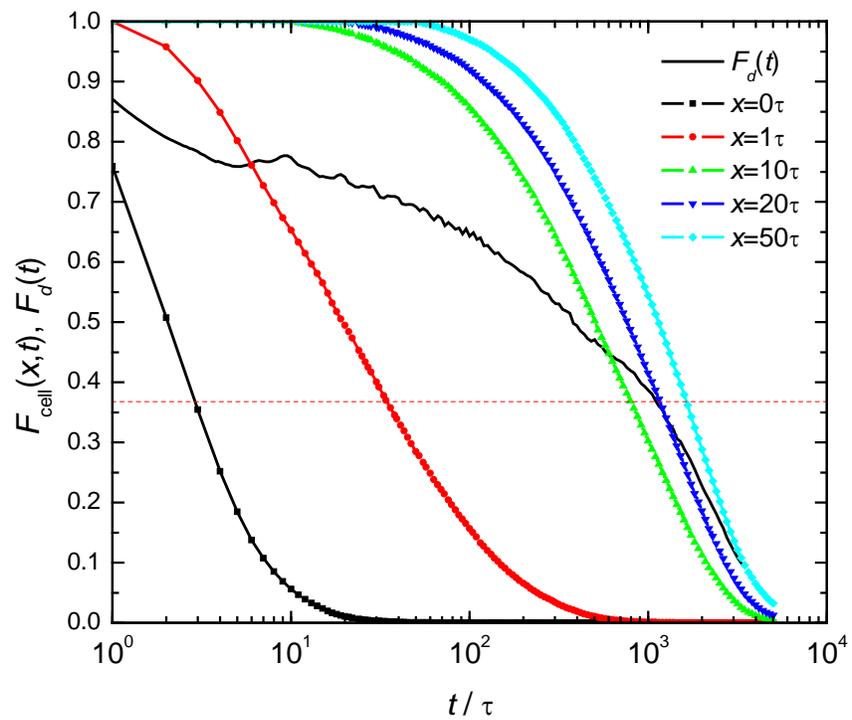

Figure 4

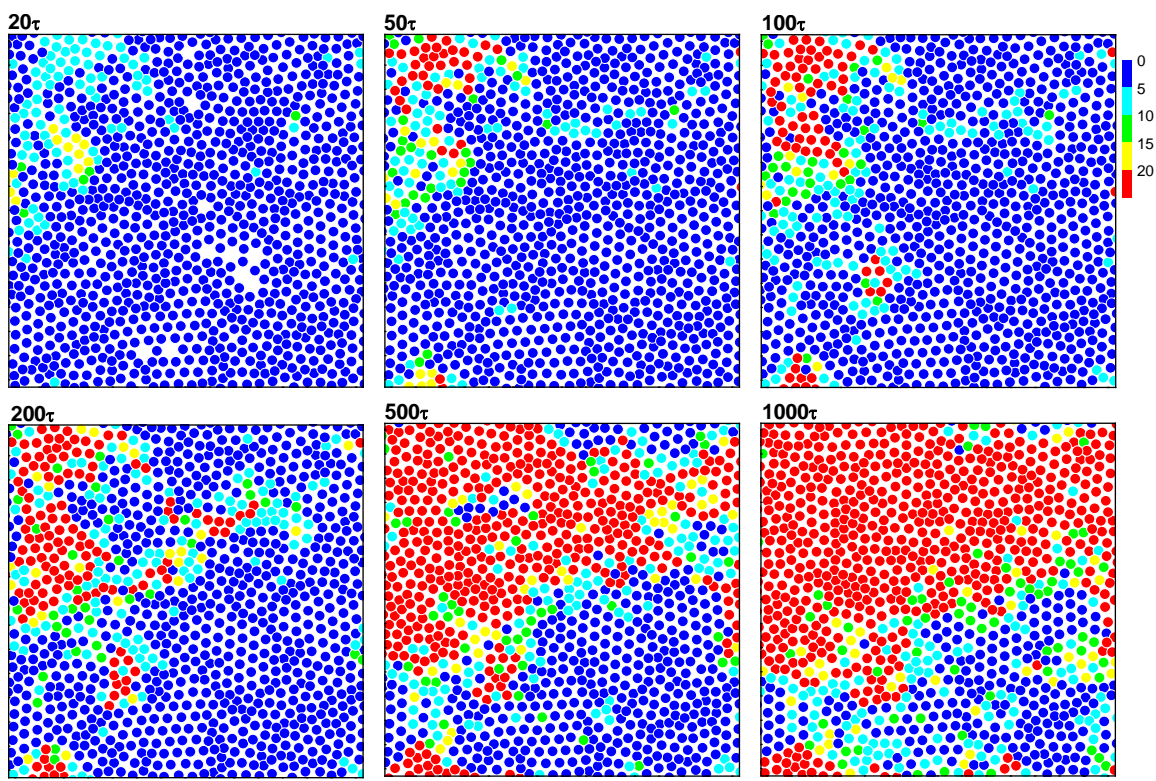

Figure 5

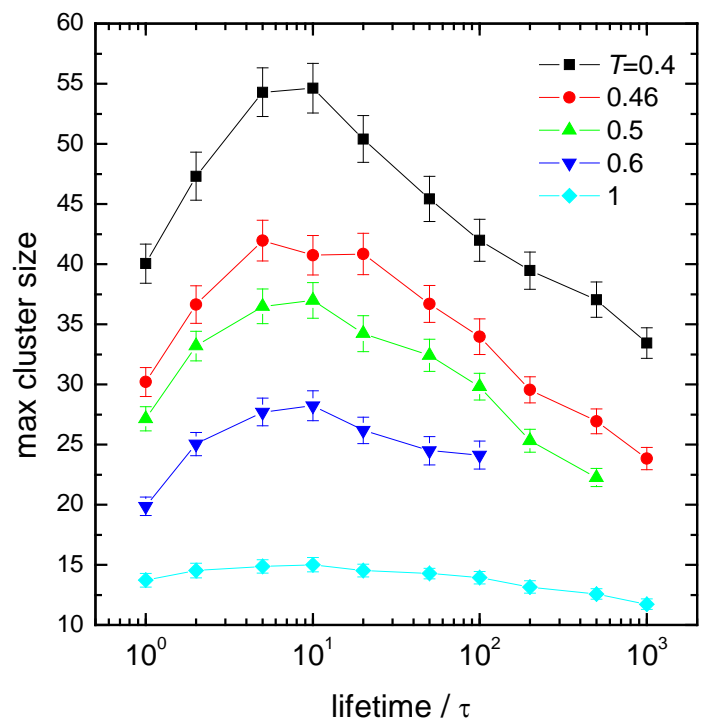

Figure 6

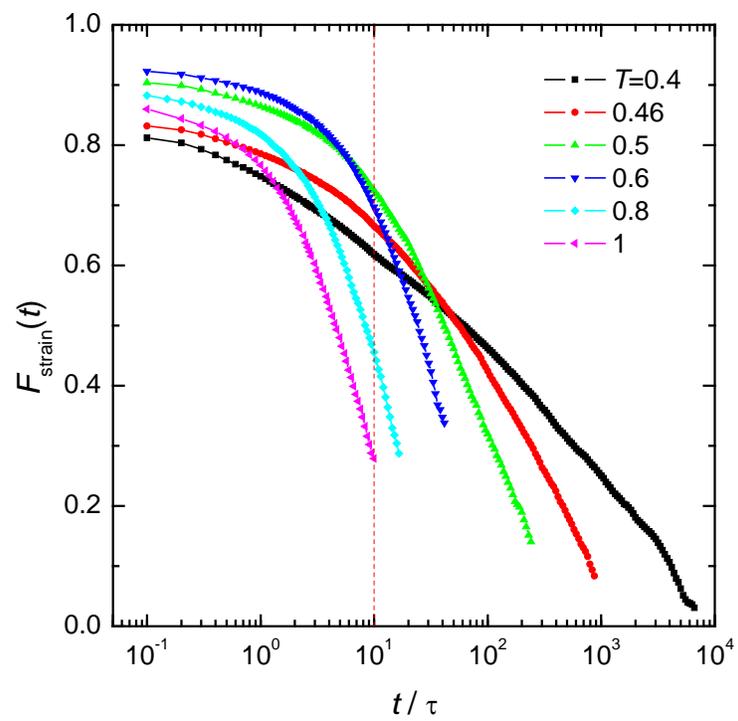

Figure 7

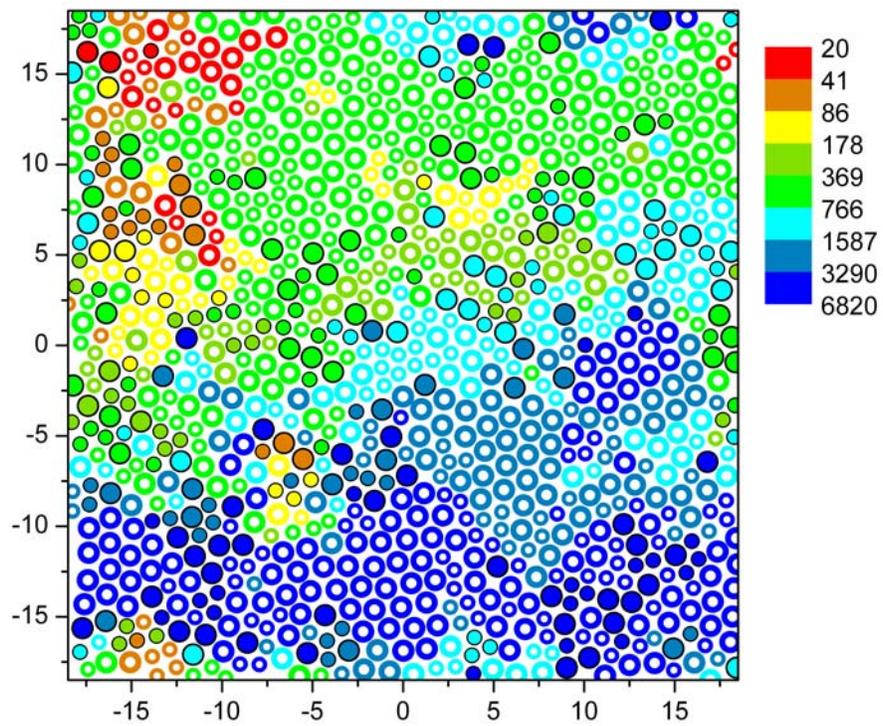

Figure 8

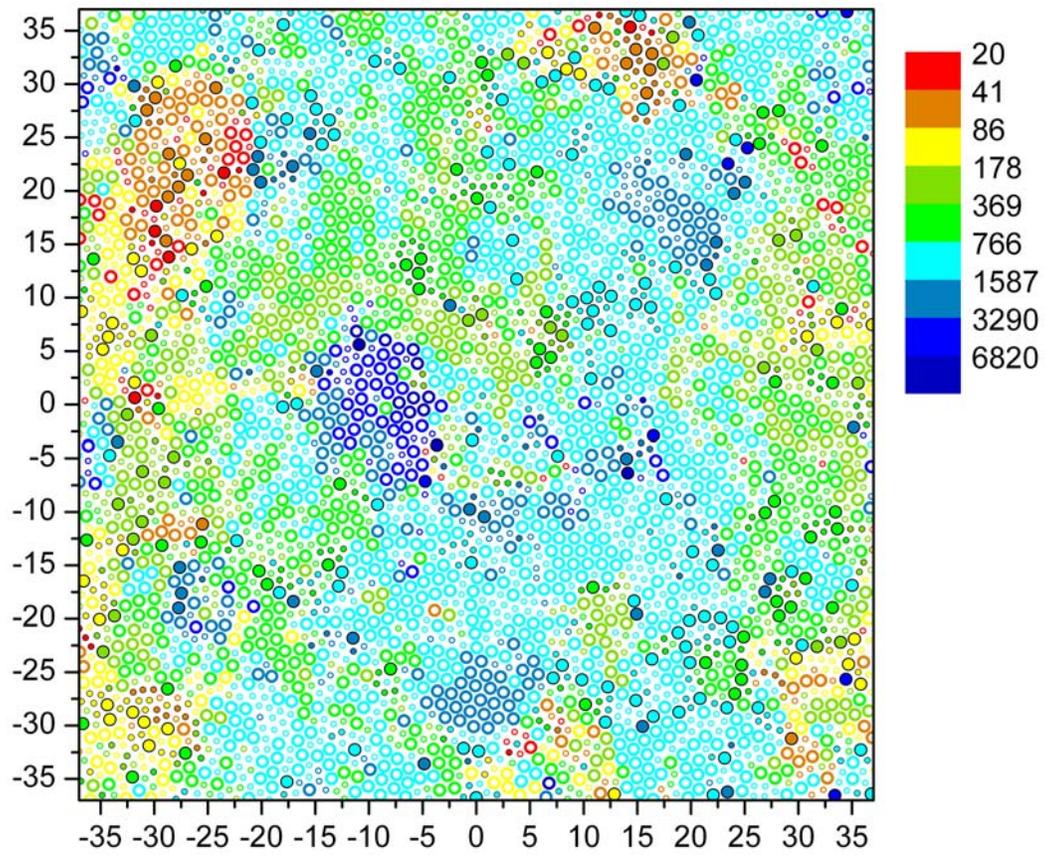

Figure 9